# Generation of High-order Group-velocity-locked Vector Solitons

**X. X. Jin,[1] Z. C. Wu,[2] Q. Zhang,[1] L. Li,[1] D. Y. Tang,[1] D. Y. Shen,[1] S. N. Fu,[2] D. M. Liu,[2] and L. M. Zhao,[1]***

[1]Jiangsu Key Laboratory of Advanced Laser Materials and Devices, School of Physics and Electronic Engineering, Jiangsu Normal University, Xuzhou 221116, China

[2]Next Generation Internet Access National Engineering Lab (NGIA), School of optical and electronic information, Huazhong University of Sci&Tech (HUST), 1037 Luoyu Road, Wuhan 430074, China

**Abstract:** We report numerical simulations on the high-order group-velocity-locked vector soliton (GVLVS) generation based on the fundamental GVLVS. The high-order GVLVS generated is characterized with a two-humped pulse along one polarization while a single-humped pulse along the orthogonal polarization. The phase difference between the two humps could be 180°. It is found that by appropriate setting the time separation between the two components of the fundamental GVLVS, the high-order GVLVS with different pulse width and pulse intensity could be obtained. “1+2” and “2+2” type high-order GVLVS could be either obtained.

## 1.Introduction

Vector solitons refer to solitons with multiple components trapped together and propagating with same group velocity in the media. Generally speaking, a SMF has weak birefringence so that there are two orthogonal polarization directions in a fiber, which provides the possibility for soliton trapping in optical fibers. Curtis R. Menyuk firstly theoretically predicted the existence of vector solitons in fibers [1], [2]. Lately, it is found that depending on the strength of fiber birefringence, various types of vector solitons, such as the phase-locked vector solitons (PLVSs) [3], the polarization rotating vector solitons [4], and the group velocity locked vector solitons (GVLVSs) [1], can be generated in a SMF. Vector solitons have also been experimentally demonstrated in mode-locked fiber lasers. Solitons generated in fiber lasers are different from those formed in fibers, which are determined by the mutual interaction between group velocity dispersion (GVD) and nonlinear Kerr effect. To investigate the soliton generation in fiber lasers we have to consider the cavity gain and loss, as well as the cavity boundary condition too. For PLVSs, their temporal and polarization state profiles during propagation maintain stationary [5], [6]. Either fundamental or high-order forms of the PLVSs were observed experimentally [7], [8]. As to GVLVSs, the orthogonal polarization components shifted their central frequencies in opposite directions through the self-phase modulation and cross-phase modulation, so regardless of the intrinsic group velocity difference caused by the fiber birefringence, the two solitons formed along orthogonal polarizations can trap each other and propagate as a non-dispersive unit [9]. The fundamental form of the GVLVS was also experimentally demonstrated [9]-[11]. However, to the best of our knowledge, no high-order GVLVS has been generated.

The extracavity vector soliton generation has also been experimentally demonstrated, in which the vector solitons are generated based on the seeding of scalar solitons obtained from a fiber laser rather than directly generated from the fiber laser [12],[13]. In this paper, we carry out the numerical simulations on an approach to generate a high-order GVLVS based on the fundamental GVLVS. By passing the fundamental GVLVS through a polarization controller with appropriate phase retard, it is feasible to obtain a high-order GVLVS which has a polarization component consisting of two humps with 180° phase difference. Depending on the time separation between the two components of the fundamental GVLVS, either “1+2” or “2+2” type high-order GVLVS could be obtained. As the high-order GVLVS is not directly generated from the fiber laser, we can call it “pseudo-high-order” GVLVS.

## 2. Theoretical Model and Simulation Results

As shown in Fig. 1, a general GVLVS from a fiber laser has two wavelength-shifted, single-humped solitons along different polarizations. A former research from our group has demonstrated the dependence of soliton central frequency difference between the two orthogonal polarization components on the cavity birefringence [9]. The small cavity birefringence leads to the solitons formed along the two orthogonal polarization directions having only slight central wavelength shift. The obvious central wavelength shift between the solitons can be obtained as the cavity birefringence becomes large. By rotating the PC, we can simultaneously change the orientation of the GVLVS and the phase difference between the orthogonal polarization components. After traversing the PC, The two polarization components of the GVLVS can be characterized by

$$F_1 = A_1 \operatorname{sech}(1.763(t - \Delta T/2)/T_0)\exp[i(2\pi ct/\lambda_1)].$$

$$F_2 = A_2 \operatorname{sech}(1.763(t + \Delta T/2)/T_0)\exp[i(2\pi ct/\lambda_2 + \varphi_t)].$$

where we assume the polarization components of the GVLVS both have $\mathrm{Sech}^2$ profile. $A_1$ and $A_2$ are the amplitudes. $\Delta T$ is the time separation between the two components. $\lambda_1$ and $\lambda_2$ are the central wavelength of the optical pulses along the two orthogonal polarizations. $\varphi_t = \varphi_i + \Delta\varphi$, in which $\varphi_i$ is the initial phase difference and $\Delta\varphi$ represents the phase difference caused by the PC. An inline polarization beam splitter (PBS) is connected to the PC. The fundamental GVLVS would further traverse the inline PBS. The two orthogonal components of the fundamental GVLVS would then be projected to the horizontal and vertical polarizations of the inline PBS and finally a high order GVLVS generated, as shown in Fig. 1. The angle between $F_1$ and the horizontal axis is $\theta$.

**2.1Status of Same Central Wavelength**

Firstly, we consider that the laser cavity birefringence is small. In this situation, the two orthogonal components of the GVLVS have nearly the same center wavelength. We use the following parameters which are consist with that obtained with small birefringence in reference [9]: $A_1 = 1$, $A_2 = 2$, $\Delta T = 0.8$ ps, $T_0 = 0.553$ ps, $\lambda_1 = \lambda_2 = 1557$ nm. Numerically we found that when $\theta = 0°$, regardless of the value of $\varphi_t$, the calculated pulse intensity profiles along the horizontal and vertical axis are both single-humped and the spectral components along the horizontal and vertical axis have the same central wavelength [Fig. 2(a)]. In this case, the two orthogonal polarization components of the vector soliton are completely resolved to horizontal and vertical axis after the PBS. However when $\theta \neq 0°$, specta dip appear at the spectrum of the two axes and the position of the spectral dip could be shifted by changing the value of $\varphi_t$ [Fig. 2(b)]. Based on the pulse intensity profiles numerically calculated, obviously the spectral dip are formed due to the spectral interference between the two humps. It is found that when $\theta = 63.4°$, $\varphi_t = 0°$, a vector soliton with a two-humped pulse along one polarization while a single-humped pulse along the orthogonal polarization could be obtained. We note that the single-humped pulse is actually superimposed by two pulses with same phase. The strong dip at the center of the spectrum of the vertical axis indicates that the phase difference between the two humps is 180° [Fig. 2(c)]. The spectrum of the two axes still present the same central wavelength. The state generated seems to be a high order GVLVS. However, it is not directly generated from the fiber laser but by tuning external PC before the external PBS. Therefore we call it a pseudo-high-order GVLVS.

Numerous simulations were further carried out to investigate the impact of $\Delta T$ on the characteristics of an obtained pseudo-high-order GVLVS. Figure 3 (a) and (b) show the pulse intensity profiles numerically obtained with the same parameters as that of Fig. 2(c) except the time separation. The time separation is chosen in intervals of 0.1 ps, from 0 to 2.5 ps. Increasing the time separation leads to a larger pulse peak intensity of the vertical component while a weaker pulse peak intensity of the horizontal component. It is found that when $0 < \Delta T < 1.1\ \mathrm{ps}$, the horizontal polarization has a single-humped pulse while the vertical polarization has a two-humped pulse. We call it a “1+2” type high-order GVLVS. As the time separation keeps growing, a small pulse begin to separate from the original pulse along the horizontal polarization. When the time separation becomes larger than 2 ps, the small pulse and the original pulse along the horizontal polarization will be completely separated. At this time, a pseudo-high-order GVLVS with a two-humped pulse along both the two orthogonal polarizations could be obtained. However, the two humps along horizontal polarization have the same phase while along the vertical polarization the two humps have 180° phase difference. We call it a “2+2” type high-order GVLVS. It is also found that the pulse width of the obtained pseudo-high-order GVLVS varies with the time separation. The pulse width was directly obtained from the pulse intensity profile calculated in our simulations with the choice of full-width-at-half-maximum (FWHM). Figure 3(c) shows the calculated pulse width versus time separation which ranges from 0 to 1.1 ps. It is clear that with time separation increasing, the pulse width of the horizontal polarization gets wider at first and then gets narrower while the pulse width of the vertical polarization gets wider monotonically. However, when the “2+2” type high-order GVLVS obtained, the generated two humps along one axis are completely separated. The overlapping between the two humps will not affect the measurement of the width of the humps. Therefore, the width of the two humps along either horizontal or vertical axis remains 0.553 ps, which is in accordance with the pulse duration of the fundamental GVLVS.

**2.2 Status of Different Central Wavelength**

If the cavity birefringence is large, the central wavelength shift between the two orthogonal components of the fundamental GVLVS will be obvious. Numerous simulations for this situation were carried out. We use the following parameters which are consist with that obtained with moderate birefringence in reference [9]: $A_1 = 1$, $A_2 = 2$ , $\Delta T = 0.8$ ps , $T_0 = 0.553$ ps , $\lambda_1 = 1557$ nm , $\lambda_2 = 1558$ nm . When $\theta = 0$, regardless of the value of $\varphi_t$, the obtained pulse intensity profiles along the horizontal and vertical axis are both single-humped pulse and the spectral components along the two axes present different central wavelength [Fig. 4(a)]. When $\theta \neq 0°$, the spectral dip whose position can be shifted by changing the value of $\varphi_t$ appeared at the spectrum. As shown in Fig. 4(b), we found that when $\theta = 63.4°$, $\varphi_t = 0°$, a "1+2" type high-order GVLVS could be obtained and the spectrum along the two axes still present wavelength shift. Compared with Fig. 2(c) in which the spectrum of vertical axis shows more than one strong spectral dip, only one strong spectral dip at he center of the spectrum of vertical axis is presented in Fig. 4(b). We note that it is caused by the different central wavelength between the two humps reducing the interference effect.

As to the different central wavelength status, we also numerically investigate the impact of $\Delta$T on the characteristics of an obtained pseudo-high-order GVLVS. We use the same parameters as that of Fig. 4(b) except the time separation. The time separation is chosen in intervals of 0.1 ps, from 0 to 2.5 ps. It is found that in the case of different central wavelength, pseudo-high-order GVLVS can be obtained even when there is no time separation between the two polarization components of the fundamental GVLVS ($\Delta$T = 0 ps), as shown in Fig. 4(c). The pulse intensity profile of the vertical axis has been multiplied by 10 times for a better display. We note that in the same center wavelength situation, in order to generate the pseudo-high-order GVLVS, there must be a time separation between the two orthogonal components. Further increasing the time separation, as shown in Fig. 5(a) and (b), we found that the pulse intensity profiles and the pulse peak intensity of the pseudo-high-order GVLVS have the similar evolutionary process as that of the same central wavelength situation. A small pulse will be separated from the original pulse along the horizontal axis as the time separation increasing and finally the "1+2" type high-order GVLVS evolves to a "2+2" type high-order GVLVS. However, compared with Fig. 3(c), the pulse width varies differently with the time separation in Fig. 5(c). With the time separation increasing, the pulse width of the horizontal component in both Fig. 3(c) and Fig. 5(c) get wider at first and then get narrower. The pulse width of the vertical components in Fig. 3(c) always gets wider while in Fig. 5(c) it gets narrower at first and then gets wider. When the time separation is big enough so that a "2+2" type high-order GVLVS generated, the width of all the humps are measured to be 0.553 ps.

We have to clarify that for the case of same central wavelength, the variation of the pulse duration depends on the pulse separation only, which means that the trends observed in Fig.3(c) are of universal nature. For the case of different central wavelength, the trend observed in Fig.5(c) only applies to the specified simulation parameters.

## 3 Conclusion

In conclusion we propose and numerically demonstrate the generation of high-order GVLVS based on the fundamental GVLVS. It is feasible to transform the fundamental GVLVS into a high-order one by passing the fundamental GVLVS through an external PC together with an inline PBS. The high-order GVLVS is characterized by a spectral dip along one polarization. Depending on the time separation between the two orthogonal polarization components of the foudamental GVLVS, “1+2” and “2+2” type high-order GVLVS could be either obtained. By appropriate setting the time separation, the high-order GVLVS with different pulse intensity and pulse width can be obtained. As it is not directly generated from the fiber laser, we consider it as the so-called ”pseudo-high-order” GVLVS.

This work was supported in part by the National Natural Science Foundation of China (61275109, 61275069, 61331010), in part by the Priority Academic Program Development of Jiangsu higher education institutions (PAPD), in part by the Jiangsu Province Science Foundation (BK20140231), in part by the National Key Scientific Instrument and Equipment Development Project (No. 2013YQ16048702) and in part by a project funded by the Jiangsu Normal University for graduate students in research and innovation (KYLX_1427).

## Reference

[1] C. R. Menyuk , “Stability of solitons in birefringent optical fibers. I: Equal propagation amplitudes,” Opt. Lett., vol. 12, no. 17, pp. 614-616, 1987.

[2] C. R. Menyuk , “Stability of solitons in birefringent optical fibers. II. Arbitrary amplitudes,” J. Opt. Soc. Amer. B, Opt. Phys., vol. 5, no. 2, pp. 392-402, 1988.

[3] D. N. Christodoulides and R. I. Joseph, “Vector solitons in birefringent nonlinear dispersive media” Opt. Lett., vol. 13, no. 1, pp. 53-55, 1988.

[4] K. J. Blow, N. J. Doran and D. Wood, “Trapping of energy into solitary waves in amplified nonlinear dispersive systems,” Opt. Lett., vol. 12, no. 12, pp. 1011-1013, 1987.

[5] N. Akhmediev, A. Buryak, J. M. Soto-Crespo. “Elliptically polarised solitons in birefringent optical fibers,” Opt. commun., vol.112, no. 5, pp. 278-282, 1994.

[6] N. Akhmediev, A. Buryak, J. M. Soto-Crespo, and D. R. Andersen. “Phase-locked stationary soliton states in birefringent nonlinear optical fibers,” J. Opt. Soc. Amer. B, Opt. Phys., vol. 12, no. 3, pp. 434-439, 1995.

[7] S. Cundiff, B. Collings & W. Knox, “Polarization locking in an isotropic, modelocked soliton Er/Yb fiber laser,” Opt. Exp., vol.1, no. 1, pp. 12-21, 1997.

[8] D. Y. Tang, H. Zhang, L. M. Zhao & X. Wu, “Observation of high-order polarization-locked vector solitons in a fiber laser,” Phys. Rev. Lett., vol. 101, no. 15, pp. 153904, 2008.

[9] L. M. Zhao, D. Y. Tang, H. Zhang, X. Wu and N.Xiang, “Soliton trapping in fiber lasers,” Opt. Exp., vol. 16, no. 13, pp. 9528-9533, 2008.

[10] L. M. Zhao, D. Y. Tang, X.Wu and H.Zhang, “ Dissipative soliton trapping in normal dispersion-fiber lasers,” Opt. Lett., vol. 35, no. 11, pp. 1902-1904, 2010.

[11] X. Yuan, T. Yang, J.Chen, X. He, H.Huang, S. Xu and Z. Yang, “Experimental observation of vector solitons in a highly birefringent cavity of ytterbium-doped fiber laser,” Opt. Exp., vol. 21, no. 20, pp. 23866-23872, 2013.

[12] A. E. Korolev, V. N. Nazarov, D. A. Nolan, & C. M. Truesdale, “Experimental observation of orthogonally polarized time-delayed optical soliton trapping in birefringent fibers,” Opt. Lett., vol.30, no. 3, pp. 132-134, 2005.

[13] D. Rand, I. Glesk, C. S. Brès, D. A. Nolan, X. Chen, & J. Koh, et al., “Observation of temporal vector soliton propagation and collision in birefringent fiber,” Phys. Rev. Lett., vol. 98, no. 5, pp. 053902, 2007.

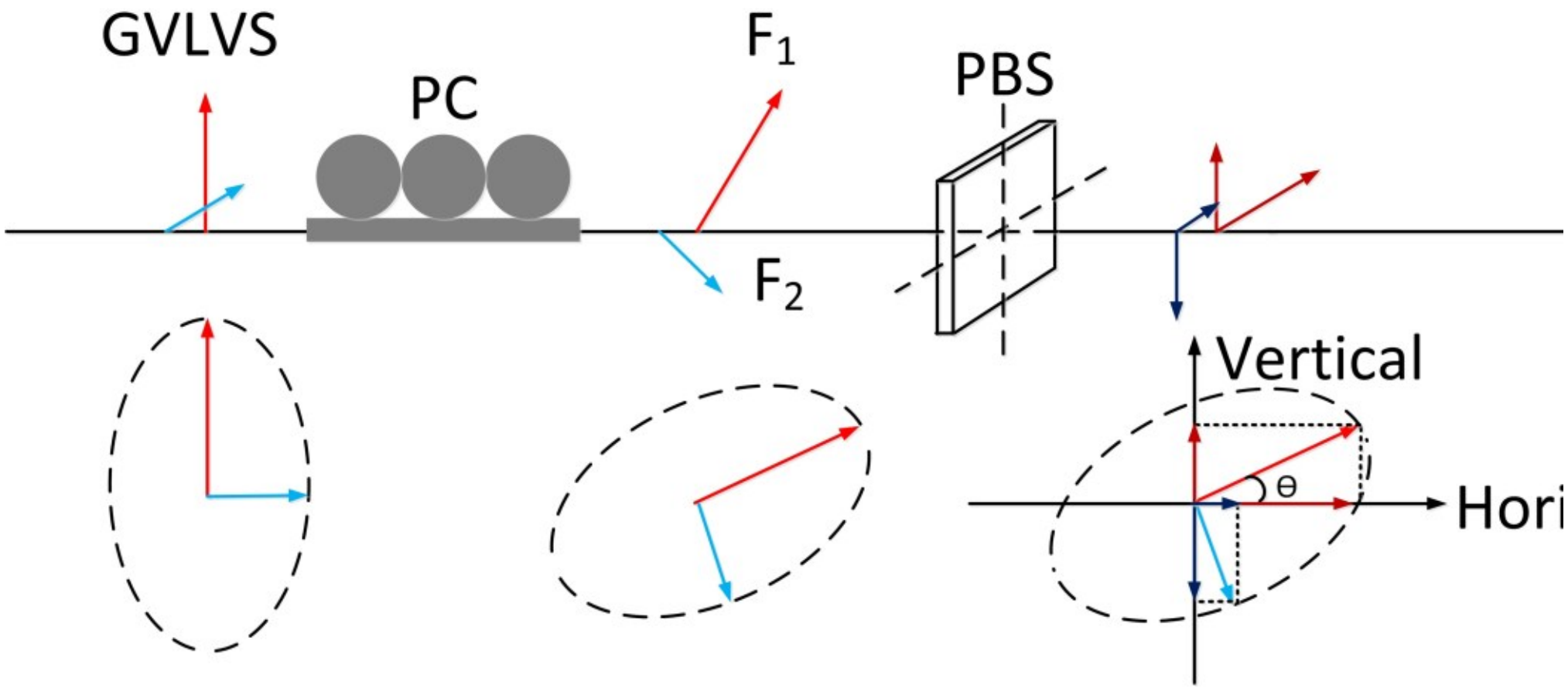

Fig. 1. Schematic of polarization evolution of a GVLVS when the input GVLVS goes through a PC and an inline PBS. PC, polarization controller; PBS, polarization beam splitter.

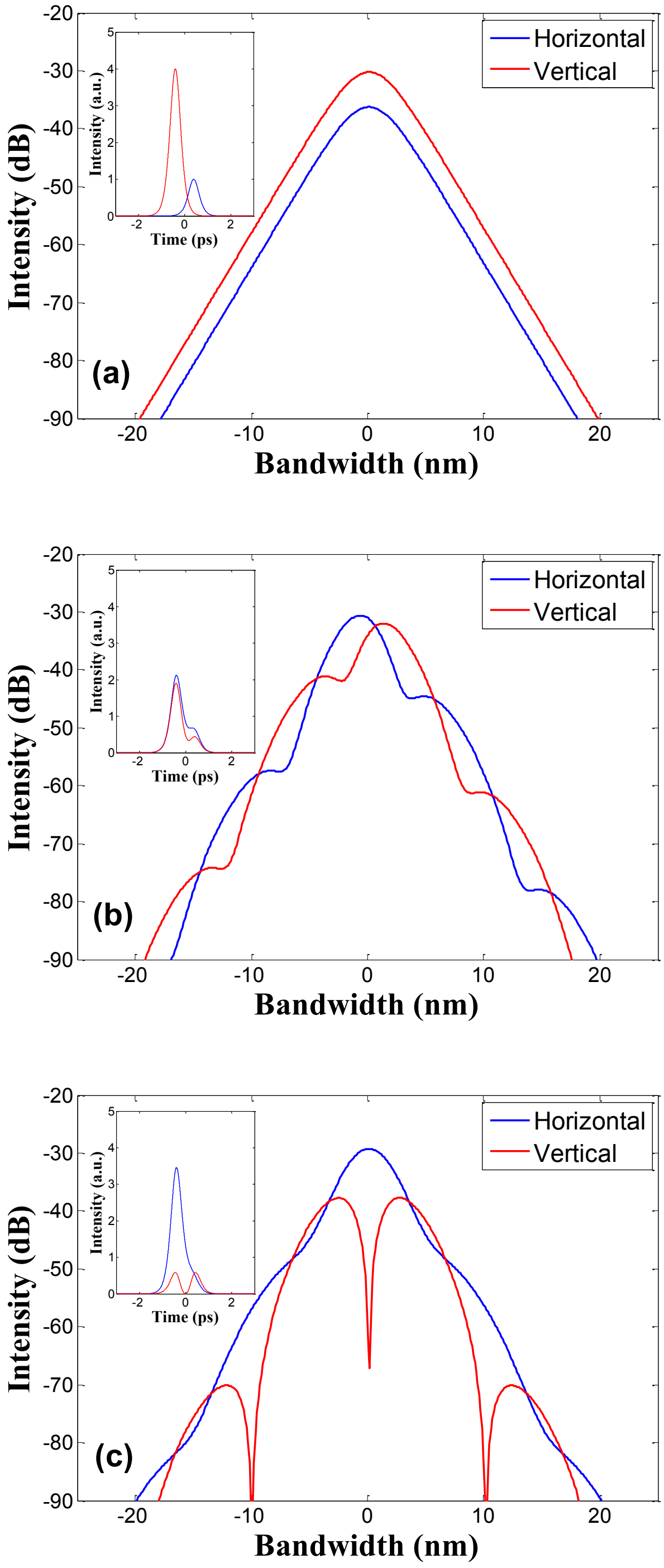

Fig. 2. Numerically obtained optical spectrum (a)$\theta = 0°$. (b)$\theta = 45°$, $\varphi_t = 70°$. (c)$\theta = 63.4°$, $\varphi_t = 0°$. Inset: corresponding temporal profile.

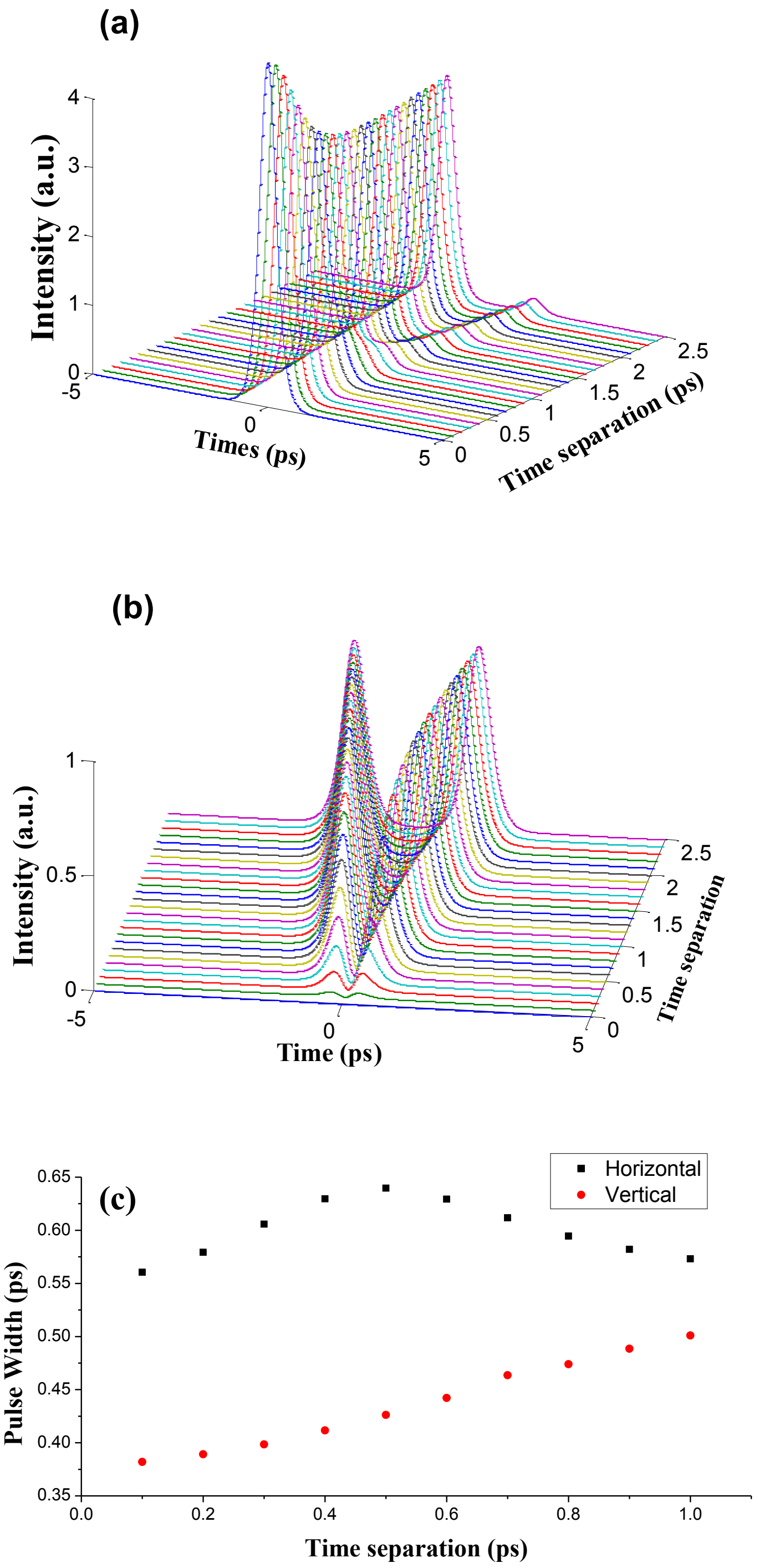

Fig. 3. Numerically obtained (a) pulse intensity profile of the horizontal axis. (b) pulse intensity profile of the vertical axis. (c) pulse width versus time separation.

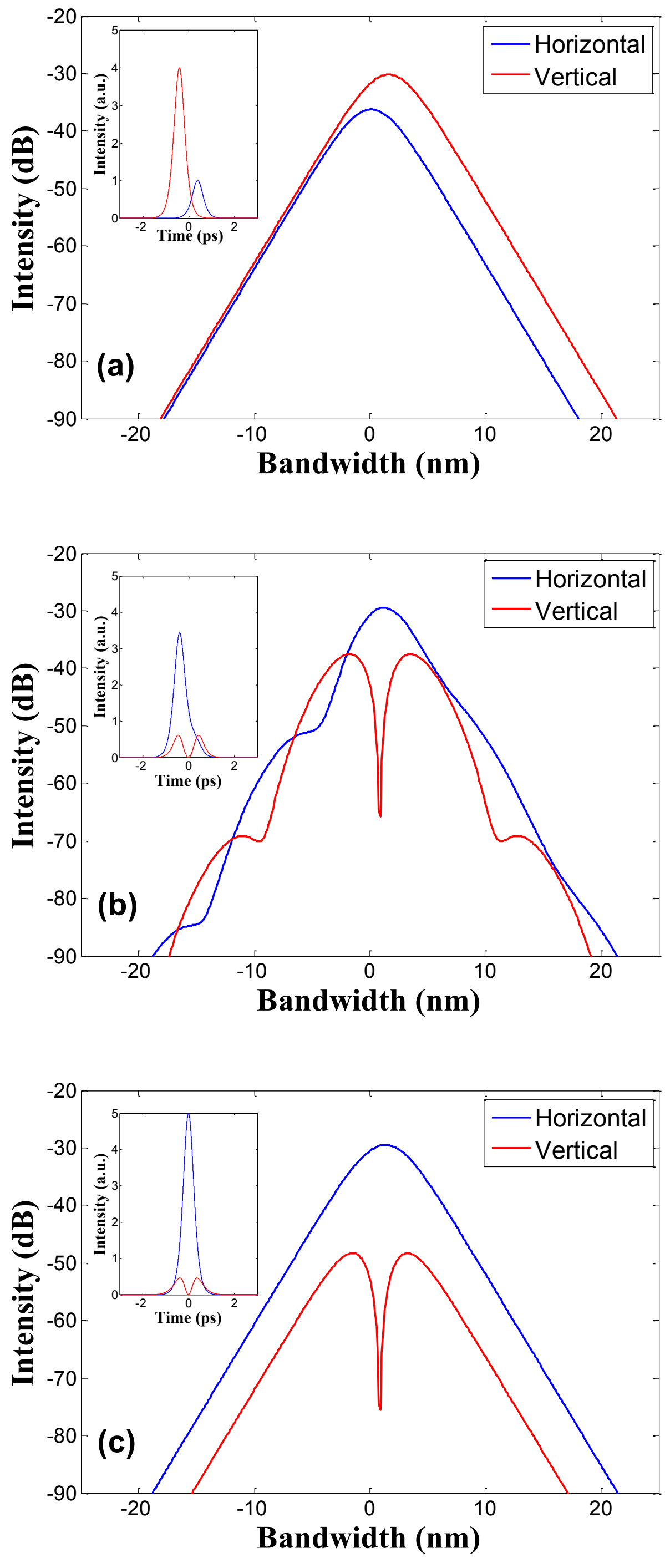

Fig. 4. Numerically obtained optical spectrum (a) $\theta = 0°$. (b) $\theta = 63.4°, \varphi_t = 0°$. (c) $\theta = 63.4°$, $\varphi_t = 0°$, $\Delta T = 0$ ps. Inset: corresponding temporal profile.

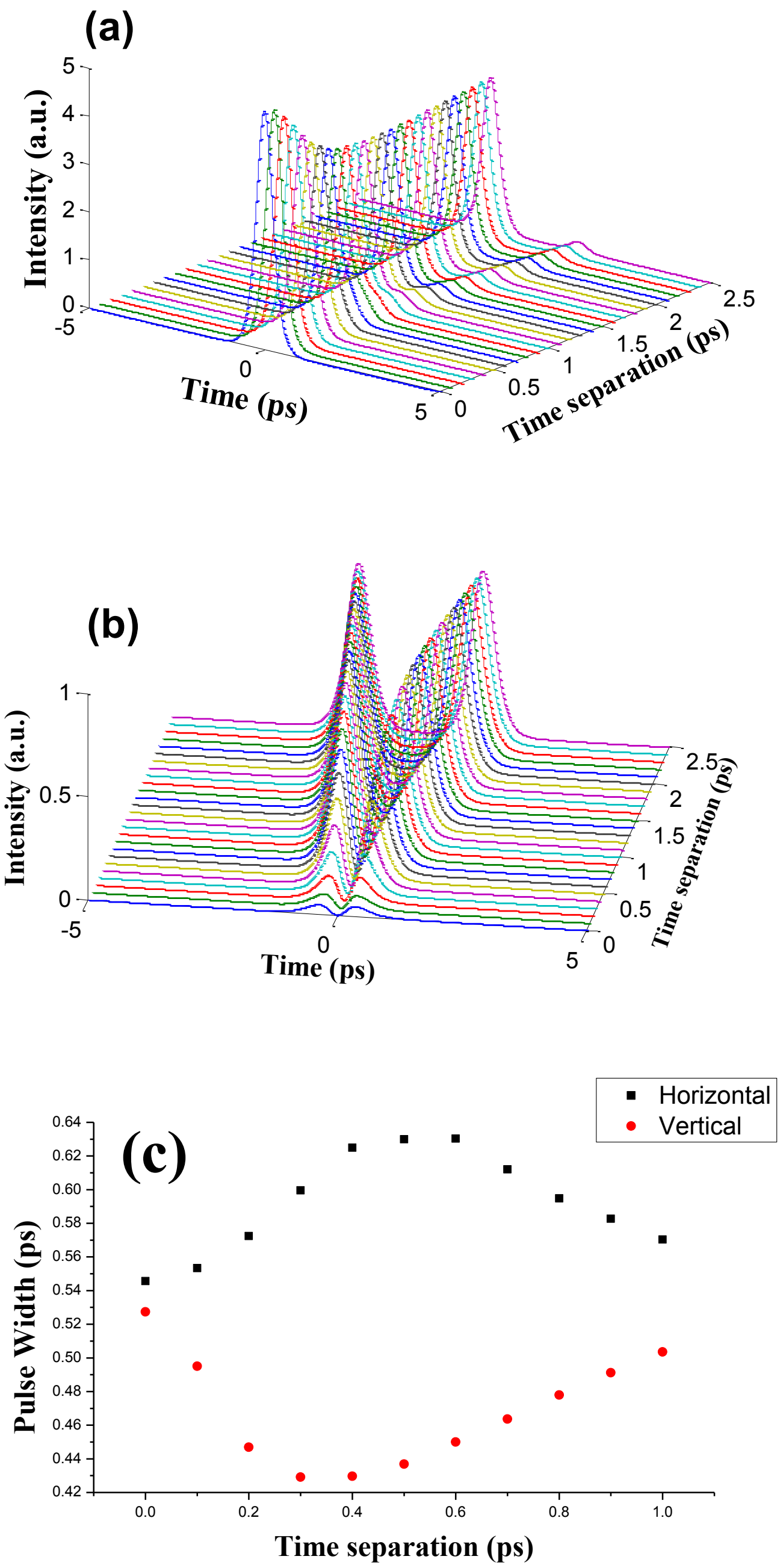

Fig. 5. Numerically obtained (a) pulse intensity profile of the vertical axis. (b) pulse intensity of the horizontal axis. (c) pulse width versus time separation.